\documentclass[aip,amsmath,amssymb,graphicx]{revtex4-1}
\usepackage{graphicx}% Include figure files
\usepackage{dcolumn}% Align table columns on decimal point
\usepackage{bm}% bold math

\draft % marks overfull lines with a black rule on the right

\begin{document}

%\preprint{}

\title{Trapped charge dynamics in InAs nanowires} %Title of paper



%\author{}
%\thanks{}
%\affiliation{}

\author{Gregory W. Holloway}
\altaffiliation{Department of Physics and Astronomy, University of Waterloo, Waterloo, Ontario, N2L 3G1, Canada}
\altaffiliation{Waterloo Institute for Nanotechnology, University of Waterloo, Waterloo, Ontario, N2L 3G1, Canada} 
\affiliation{Institute for Quantum Computing, University of Waterloo, Waterloo, Ontario, N2L 3G1, Canada}

\author{Yipu Song}
\altaffiliation{Department of Chemistry, University of Waterloo, Waterloo, Ontario, N2L 3G1, Canada}
\affiliation{Institute for Quantum Computing, University of Waterloo, Waterloo, Ontario, N2L 3G1, Canada}

\author{Chris M. Haapamaki}
\affiliation{Department of Engineering Physics, Centre for Emerging Device Technologies, McMaster University, Hamilton, ON, L8S 4L7, Canada}

\author{Ray R. LaPierre}
\affiliation{Department of Engineering Physics, Centre for Emerging Device Technologies, McMaster University, Hamilton, ON, L8S 4L7, Canada}

\author{Jonathan Baugh}
\email{baugh@iqc.ca}
\altaffiliation{Department of Chemistry, University of Waterloo, Waterloo, Ontario, N2L 3G1, Canada}
\altaffiliation{Department of Physics and Astronomy, University of Waterloo, Waterloo, Ontario, N2L 3G1, Canada}
\altaffiliation{Waterloo Institute for Nanotechnology, University of Waterloo, Waterloo, Ontario, N2L 3G1, Canada} 
\affiliation{Institute for Quantum Computing, University of Waterloo, Waterloo, Ontario, N2L 3G1, Canada}

\date{\today}

\begin{abstract}
We study random telegraph noise in the conductance of InAs nanowire field-effect transistors due to single electron trapping in defects. The electron capture and emission times are measured as functions of temperature and gate voltage for individual traps, and are consistent with traps residing in the few-nanometer-thick native oxide, with a Coulomb barrier to trapping. These results suggest that oxide removal from the nanowire surface, with proper passivation to prevent regrowth, should lead to the reduction or elimination of random telegraph noise, an important obstacle for sensitive experiments at the single electron level.
\end{abstract}

\pacs{73.63.-b}% insert suggested PACS numbers in braces on next line

\maketitle %\maketitle must follow title, authors, abstract and \pacs

\section{Introduction}
\indent InAs nanowires continue to attract much attention as an interesting material for nanoscale circuits \cite{Nam09}, spin-dependent quantum transport \cite{Nadj-Perge10}, single electron charge sensing \cite{Shorubalko2008,Salfi2010} and potentially for realizing topological quantum states \cite{Gangadharaiah2011, Mourik2012, Sau2010}. A serious impediment to obtaining clean behaviour in transport devices is the uncontrolled spatial variation of electrostatic potential along the nanowire, evidenced by spontaneous quantum dot formation at low temperatures \cite{Schroer2010}. These fluctuations may be due to surface defects \cite{Dayeh2007}, stacking faults \cite{Schroer2010}, or charged defects in the nanowire or in the native oxide layer \cite{amarasinghe01, Kirton89, Salfi2012}. Fluctuations due to charge traps can vary in time due to carrier trapping and detrapping events, leading to the appearance of random telegraph noise (RTN) in the device conductance. The large nanowire surface-to-volume ratio renders nanowire transistors very sensitive to these charge fluctuations \cite{Salfi2010, Salfi2011}. We have observed and studied RTN in a number of InAs field-effect transistor (FET) devices and here show results consistent with the charge traps giving rise to RTN residing in the oxide. The charge dynamics are consistent with a charge trap model that includes a Coulomb energy barrier \cite{schulz93} in addition to a multiphonon emission barrier \cite{Salfi2010}. These results confirm that the native oxide is the main source of charge noise in high quality InAs nanowires, and help to shed light on the underlying physics of the trapped charge dynamics. \\
\indent In order to study the trapped charge behaviour, we employ FET devices in which the global potential of the nanowire channel is adjusted using a backgate. The nominally undoped nanowires are $n$-type, due to the presence of surface states which act as electron donors and give rise to a surface accumulation layer \cite{Dayeh2007, Noguchi1991}. At certain temperatures and at sufficiently slow gate sweep rates, random jumps can be seen in the source-drain conductance (figure 1a). These shifts are evidence of the changes in local potential that occur as the charge state of a trap changes by one electron. The trapped electron generates an electric field in the nanowire that produces a potential barrier, and local depletion of carriers, reducing conductance \cite{Kirton89,Lee2010, Salfi2012}. By setting the gate voltage to be constant near one such step and measuring the DC conductance versus time with sufficiently large bandwidth, RTN can be observed and recorded (figure 1b). Occasionally, we have seen multilevel fluctuations reflecting the dynamics of multiple traps \cite{Seungwon2008}, but here we focus on single trap behaviour. Guided by the Coulomb barrier model of Schulz \cite{schulz93}, we perform experiments in which the modulation of gate voltage and temperature are used to determine the activation energies and place upper bounds on the radial locations of individual charge traps. The capture and emission dynamics we observe are consistent with traps that are charge neutral in the empty state and negatively charged in the filled state, i.e. electron traps. \\
\begin{figure}
\includegraphics[width=1\textwidth]{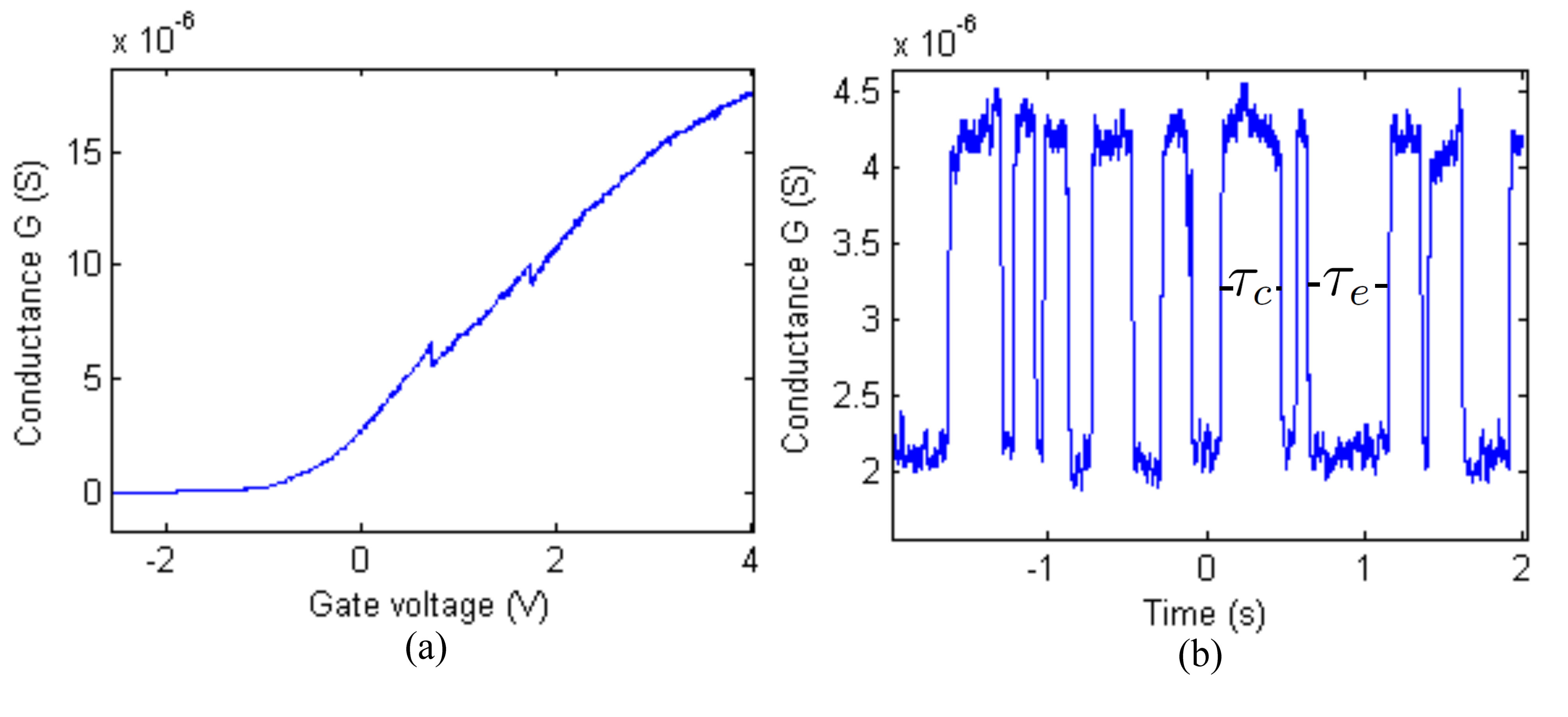}
\caption{(a) Conductance through an InAs nanowire FET as gate voltage is swept from negative to positive values. The two visible jumps are caused by electron capture events in two different charge traps. (b) Random telegraph signal in the FET conductance versus time, showing two-level behaviour. The electron capture and emission times, $\tau_c$ and $\tau_e$, correspond to the high and low conductance states, respectively.}\label{fig1}
\end{figure}
\section{Methods}
\indent The InAs nanowires used here are grown in a gas source molecular beam epitaxy system using Au seed particles \cite{Plante2009}. Nanowires are mechanically deposited onto a 180 nm thick layer of SiO$_2$ on top of a degenerately doped silicon wafer. Using scanning electron microscopy, we select untapered nanowires with diameters 30-60 nm for contacting. Ni/Au Ohmic contacts are deposited after an etching/passivation step \cite{Suyatin2007}, with a typical FET channel length of 1 $\mu$m.  The sample is then wire-bonded to a chip carrier and cooled in liquid helium vapour, with temperature controllable between 4 and 300 K. Differential conductance at low frequencies ($0.1-2$ kHz) is measured with a standard lock-in and current-voltage preamplifier circuit. The lock-in output is measured with a digital oscilloscope, and conductance traces up to 20 s long are recorded. The measurement bandwidth is determined by the filter of the lock-in, and for these experiments was in the range of $0.3-3$ kHz. \\
\section{Trapped charge dynamics}
\indent To study the trap dynamics we measure the average capture and emission times of individual traps. Because RTN is known to follow Poisson statistics \cite{Kirton89}, these times are obtained by taking an average over many conductance jumps. The capture and emission times can be described by the Shockley-Read-Hall relations as \cite{schulz93}:
\begin{align}
\langle \tau_c \rangle = 1/(nC_n) =\left[N_CC_n e^{-(E_C-E_F)/{k_BT}}\right]^{-1}\label{eq1}\\
\langle \tau_e \rangle = \left[N_CC_n e^{-(E_C-E_T)/{k_BT}}\right]^{-1},\label{eq2}
\end{align}
where the average capture time $\langle \tau_c \rangle$ is inversely related to the product of the density of free electrons $n$ and the capture coefficient $C_n$. The density of electrons can also be expressed through Boltzmann statistics using the energy difference between the conduction band energy $E_C$ and Fermi level $E_F$, where $N_C$ is the effective density of states in the conduction band, $k_B$ is the Boltzmann constant and $T$ is the temperature. Similarly, the emission time reflects the energy separation between the conduction band and the trap energy level $E_T$. \\
\begin{figure}
\includegraphics[width=1\textwidth]{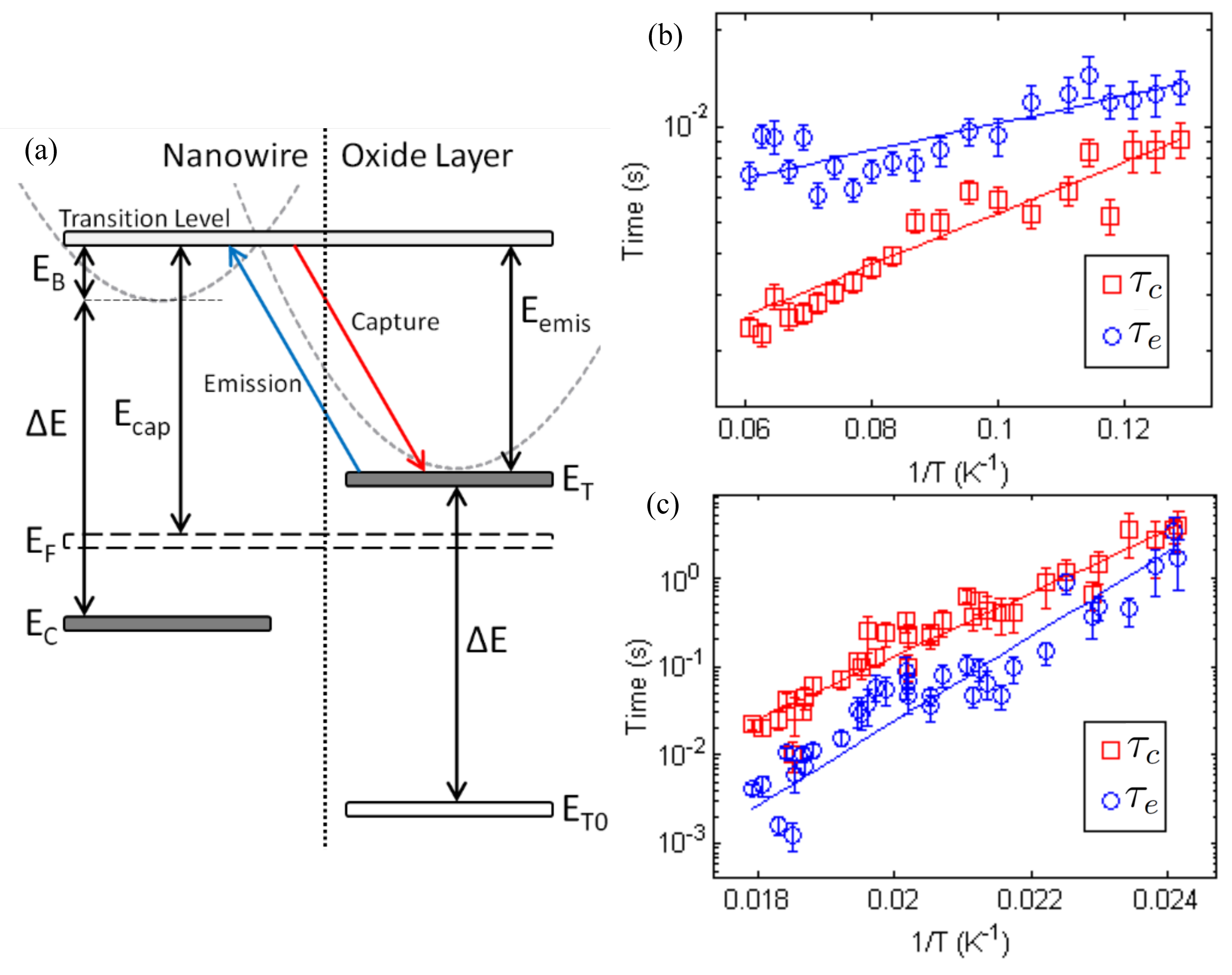}
\caption{(a) Energy level diagram describing a trap model consistent with our data. $E_F$ and $E_C$ are energies of the Fermi level and the conduction band in the nanowire. The vertical dotted line separates the nanowire and its native oxide. The dashed parabolas represent the quadratic dependence of the electron-lattice interaction energy on the configuration coordinate (not shown), which leads to the multiphonon emission barrier \cite{Salfi2010} of energy $E_B$. $E_T$ and $E_{T0}$ are the energies of the filled and empty trap states. The upper horizontal line indicates the energy of the transition level that is $\Delta{E}+E_B$ above the conduction band, where $\Delta{E}=E_T-E_{T0}$ is the Coulomb energy. $E_{cap}$ and $E_{emis}$ are the energies required for electron capture and emission to occur. $E_{cap}$ varies linearly with $\Delta{E}$, whereas $E_{emis}$ is independent of $\Delta{E}$. (b,c) Variation in the average capture and emission times of two different traps in the same FET device, versus $1/T$. The fits (solid lines) described in the text yield the energy barriers associated with capture and emission.}
\end{figure}
%%%%%%%%%
\indent The typical conduction electron density in the nanowire is $\sim 10^{17}-10^{18}$ cm$^{-3}$, including a surface accumulation layer, suggesting that the capture time should be short for a trap in the nanowire or on its surface. Experimentally, however, we find that the capture time can often be on the order of seconds or longer. This discrepancy suggests two things: (i) the capture coefficient must be small, indicating a trap located outside the conduction volume (e.g. in the native oxide), and (ii) there is an additional energy barrier that must be overcome to change the trap occupancy. For a trap located in an insulating region adjacent to a semiconductor populated with carriers, there is a Coulomb energy associated with the image charge that is created when an electron transfers from the conduction band to the trap, i.e. the trap may be modelled as a capacitor with a corresponding charging energy. For traps only a few nanometers from the semiconductor surface, this charging energy is typically on the order of 100 meV \cite{schulz93}, which leads to a large deviation of capture and emission times from the Shockley-Read-Hall predictions. It can be written $\Delta{E} = (qx_T/T_{ox})(V_G-V_{FB}-\Psi_S)$ \cite{schulz93, amarasinghe01}, where $q$ is the trapped charge, $x_T$ is the trap location relative to the nanowire surface, $T_{ox}$ is the thickness of the gate oxide, $V_G$ is the back gate voltage, $V_{FB}$ is the flat-band voltage, and $\Psi_S$ is the surface potential. Taking into account this Coulomb energy, we replace $E_C-E_F\rightarrow E_C-E_F+\Delta{E}$ in equation ~\ref{eq1} and $E_T\rightarrow E_{T0}$ in equation ~\ref{eq2}, where $E_{T0}$ is the energy level of the empty trap. It is also necessary to include an energy $E_B$ in both equations ~\ref{eq1} and ~\ref{eq2} to account for a multiphonon emission process \cite{Salfi2011, Lu2005}. This term is gate voltage independent and is the energy barrier for the simultaneous emission of several optical phonons. This process mediates the transition of the electron-lattice configuration coordinate between the free and bound electron states \cite{Shinozuka1993}.\\
\indent The energy barriers for capture and emission can now be written as $E_{cap} = E_C-E_F+\Delta{E}+E_B$ and $E_{emis} = E_C-E_{T0}+E_B$. The corresponding energy level diagram is shown in figure 2a. Trapping occurs when an electron at the Fermi level gains sufficient energy to reach the transition energy level, from which it can enter the trap at energy $E_T$. Emission occurs when a trapped electron can overcome the energy barrier $E_{emis}$. Both processes are thermally activated. Gate-induced changes in $\Delta{E}$ will cause the transition level and $E_T$ to shift together relative to the other levels. The difference between them is constant and equal to $E_{emis}$. Conversely, $E_{cap}$ depends on the separation between $E_F$ and the transition level and varies linearly with $\Delta{E}$. Both of these predictions are consistent with the RTN data shown in figure 3. The small gate voltage dependence for $E_{emis}$ seen in figure 3 can be explained by a weak dependence of $E_C-E_F$ with gate voltage. The energy level diagram in figure 2a and the expressions for $\langle \tau_c \rangle$ and $\langle \tau_e \rangle$ are consistent with the data shown in figures 2b, 2c and 3, and correspond to a trap charge state that is neutral when empty and negative when filled \cite{schulz93}. For a positive/neutral charge state, varying the gate voltage should lead to capture and emission times changing at the same rate \cite{schulz93}. This adds support to the identification of these defects as oxide charge traps, as the InAs surface donor-like states are expected to have positive/neutral charge states \cite{Dayeh2007, Salfi2012}.\\
%%%%
\indent By measuring the average capture and emission times versus changes in temperature and gate voltage, the expressions for $\langle \tau_c \rangle$ and $\langle \tau_e \rangle$ allow us to extract information on the trap energetics. The temperature dependence of capture and emission times is fit to $\langle \tau_{c,e} \rangle = \alpha e^{E_{cap,emis}/k_BT}$. From $\alpha$ we obtain $N_CC_n$ for both capture and emission. For each trap studied in detail, these coefficients were equal, within error. Additionally, this fit yields the activation energies of trapping and detrapping $E_{cap} = E_C-E_F+\Delta{E}+E_B$ and $E_{emis} = E_C-E_{T0}+E_B$. Upon studying a number of traps, a broad range of activation energies is observed. At temperatures from $8-18$ K, we find for one trap $E_{cap} = 1.6$ meV and $E_{emis} = 0.9$ meV (figure 2b). In the temperature range $40-60$ K, we find $E_{cap} = 71$ meV and $E_{emis} = 94$ meV (figure 2c) for another trap. In the $40-60$ K range, conductance jumps from the first trap are no longer observed within the bandwidth of our measurements. This is understood by considering that conductance fluctuations will occur too rapidly to be resolved when thermal energy is larger than the activation energy. On the other hand, for traps with activation energy much larger than thermal energy, electrons are unable to overcome the energy barrier, and so the charge state will be frozen in. RTN observed at higher temperatures therefore arises predominantly from traps with larger activation energies. \\
\begin{figure}
\includegraphics[width=1\textwidth]{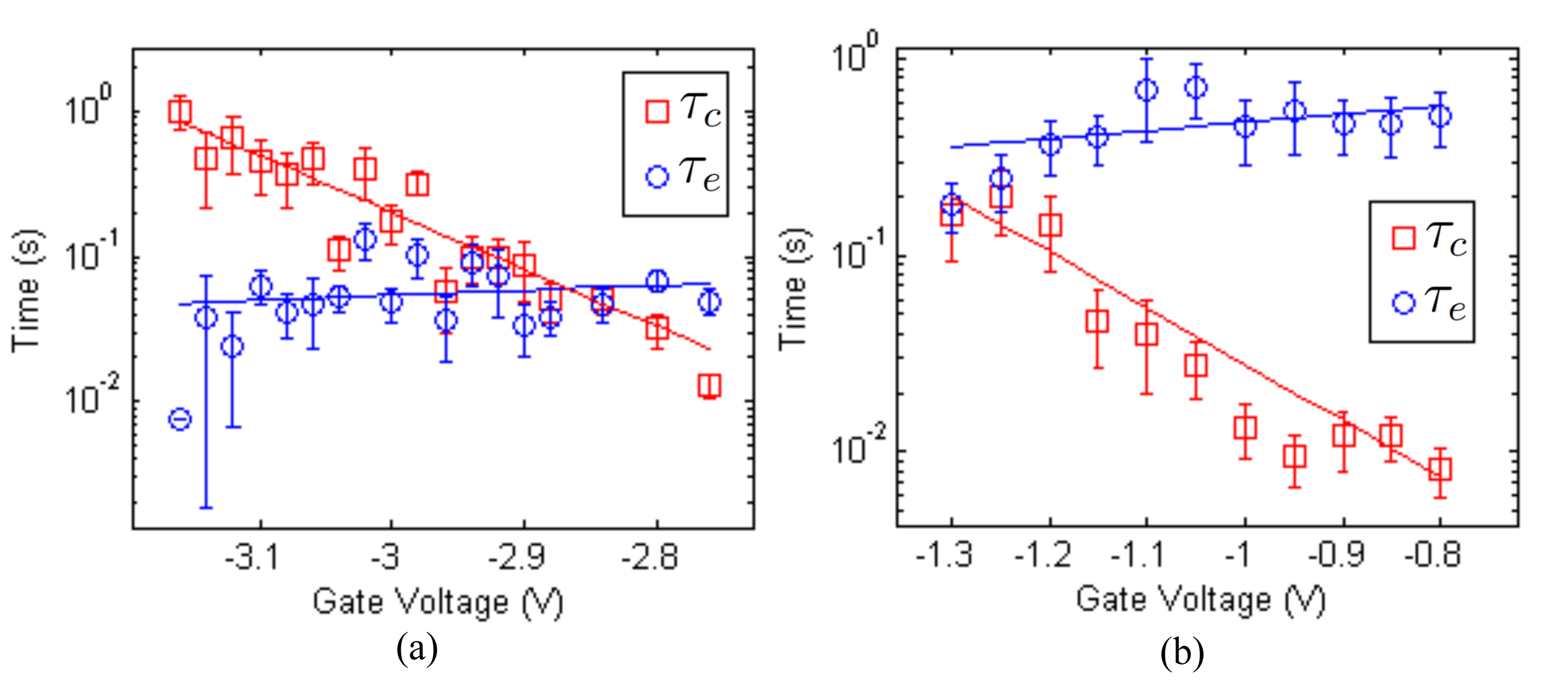}
\caption{(a,b) Average capture and emission times as a function of gate voltage $V_G$. The weak dependence of the emission time on $V_G$ is consistent with a model of neutral/negative charge traps. The dependence of $\langle\tau_c\rangle/\langle\tau_e\rangle$ on $V_G$ is used to extract an upper bound on the radial distance of the trap relative to the nanowire surface.}
\end{figure}
\indent The Fermi level being pinned above the conduction band at the nanowire surface \cite{Noguchi1991} allows us to estimate of the magnitude of $\Delta{E}$. The Fermi level is typically $0-0.26$ eV above the conduction band \cite{Affentauschegg2001}. For the trap corresponding to the data in figure 2c, where $E_{cap} = 71$ meV, the expression $E_{cap} = E_C-E_F+\Delta{E}+E_B$ suggests $\Delta{E}+E_B$ is roughly in the range $71-330$ meV. This is consistent with the expectation $\Delta{E} \sim 100$ meV noted by Schulz \cite{schulz93}. The gate voltage dependence of the capture and emission times allows us to estimate the radial distance of a trap from the nanowire surface. From the definition of $\Delta{E}$ there is an explicit dependence on gate voltage $V_G$. Therefore by fitting the ratio of capture time over emission time to $\langle \tau_c \rangle/\langle\tau_e\rangle = \beta e^{\gamma V_G}$, we may estimate the separation of the trap from the nanowire surface, $x_T = \gamma T_{ox}k_BT/q$. However, this calculation neglects the dependence of the surface potential $\Psi_S$ on gate voltage, leading to an overestimate for the value of $x_T$ \cite{Lee2007}, so we treat this estimate of $x_T$ as an upper bound. For the two traps shown in figures 3a and 3b, $x_T \leq 8.4$ $\pm$ 3.2 nm and 11.8 $\pm$ 4.0 nm, respectively, whereas the amorphous oxide layer of the nanowire is known to be approximately $2-5$ nm thick from transmission electron microscopy. For the charge traps that induced the largest conductance jumps in our experiments, we estimate a charge sensitivity \cite{Salfi2010} $\approx 6\times 10^{-4}$ $e/\sqrt{Hz}$. \\
\begin{figure}
\includegraphics[width=1\textwidth]{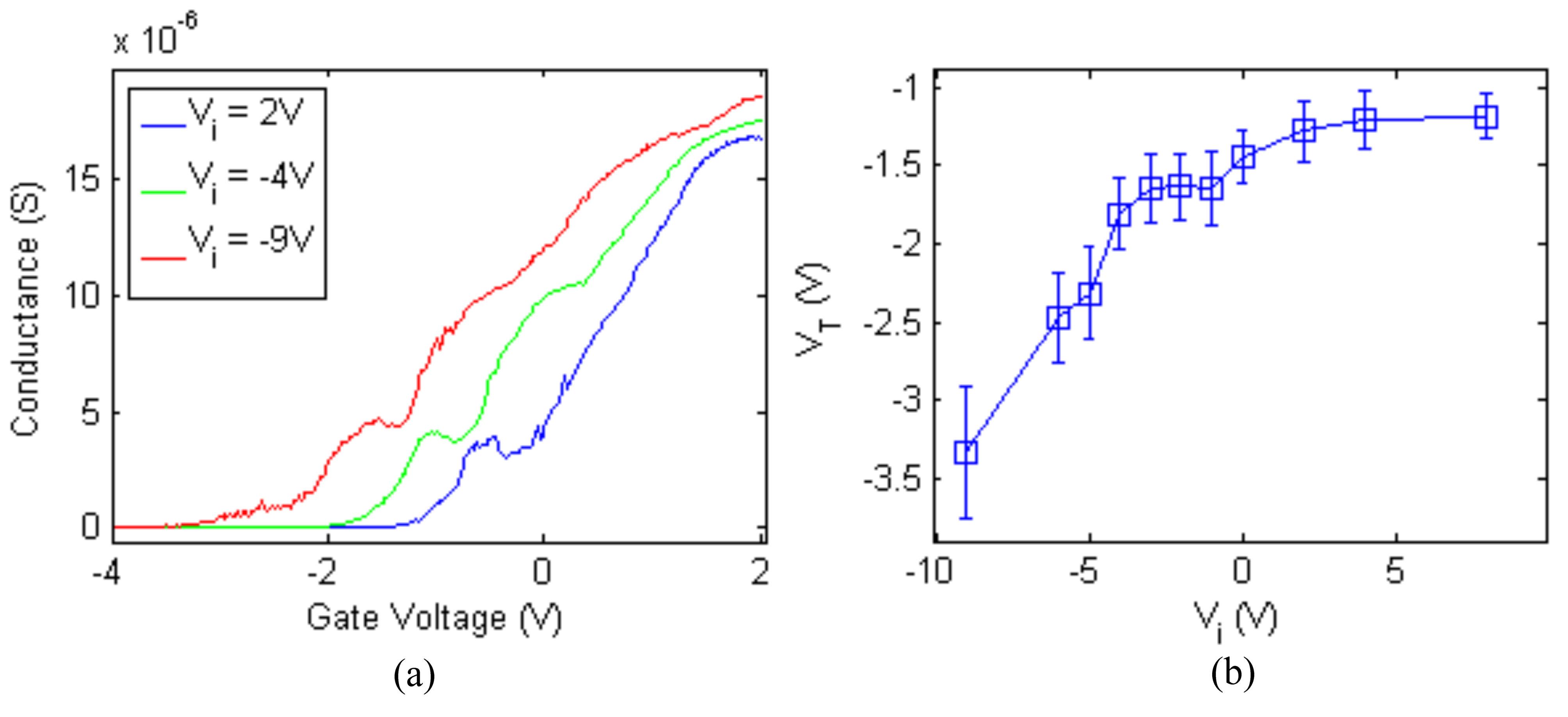}
\caption{(a) Conductance curves for a $32$ nm diameter nanowire FET measured at $T=$50 K, with several initial gate voltages $V_i$ applied during cooldown from $T>150$ K. When a positive $V_i$ is applied, traps are predominantly filled and pinchoff occurs at a more positive gate voltage. Here `traps' may refer to other defects beyond native oxide charge traps, such as InAs surface states or SiO$_2$ charge traps. (b) Change in pinchoff threshold voltage $V_T$ versus $V_i$. The saturation that occurs at positive $V_i$ suggests most traps are being filled. No saturation was seen for negative voltages down to $-9$ V, suggesting that only a fraction of traps were depleted.}
\end{figure}
\section{Hysteretic behaviour}
\indent Finally, we include data showing hysteretic effects that may arise, at least in part, from changes in the trapped charge population. An initial gate voltage $V_i$ was applied at $T > 150$ K and during cooling to freeze in a particular charge configuration. Conductance curves measured at $T = 50$ K corresponding to three $V_i$ values are shown in figure 4a for a 32 $\pm 2$ nm diameter nanowire with channel length 820 nm. As an aside, reproducible plateau-like features can clearly be seen in the conductance. These features could be due to populating quantized radial subbands of the nanowire, or due to resonant scattering from a defect. The conductance curve shifts to more negative voltages as $V_i$ is made more negative. This is consistent with the expectation that as more traps are depleted of electrons, the average conduction volume in the nanowire increases, requiring more negative gate voltage to reach pinchoff. The average slope of the conductance curve also decreases with more negative $V_i$, indicating a lower effective mobility. This is also consistent with a greater fraction of carrier density near the nanowire surface, which should dominate scattering. In particular, at $V_i=-9$ V we observe a pronounced low-mobility tail just before pinchoff. We cannot assign the shift in conductance exclusively to the nanowire oxide traps, since SiO$_2$ charge traps and the gate-induced ionization of InAs surface states may also contribute. \\
\section{Conclusion}
\indent We have shown that an oxide trap model in which electrons must overcome a Coulomb barrier, in addition to a multiphonon emission barrier, to move into a bound state correctly describes sources of RTN in InAs nanowire FET devices. The model was used to extract activation energies and to place upper bounds on the radial locations of several distinct traps. Due to the appreciable density and broad activation energy range of these oxide traps, RTN is commonly observed in nanowire electronics, and hinders the performance and stability of single-electron devices. Our results suggest that oxide removal from the nanowire surface, with proper passivation to prevent regrowth, should lead to the reduction or elimination of RTN, an important obstacle for sensitive experiments at the single electron level. Recent advances in chemical passivation \cite{sun2012} might accomplish this. Further research on epitaxial core-shell nanowires \cite{Haapamaki2012, Tilburg2010}, where the oxide is physically separated from the active channel, could lead to reduced RTN and also improve the uniformity of the electric potential along the nanowire. Despite the detrimental effects of charge traps, they are useful for assessing the charge sensitivity of nanowire transistors \cite{Salfi2010}. \\

\textbf{Acknowledgements --} 
We would like to acknowledge the Canadian Centre for Electron Microscopy, the Centre for Emerging Device Technologies, and the Quantum NanoFab facility for technical support. We thank Shahram Tavakoli for assistance with MBE and Roberto Romero for general technical assistance. We thank Joseph Salfi for helpful comments. This research was supported by NSERC, the Ontario Ministry of Economic Innovation, the Canada Foundation for Innovation and Industry Canada. G. W. H. acknowledges a WIN Fellowship.

\bibliographystyle{apsrev4-1}
\bibliography{rtn}

\end{document}